# Rapid Prototyping of Standard Compliant Visible Light Communications System

Ciprian. G. Gavrincea*, Jorge. Baranda*, and Pol. Henarejos*
*Centre Tecnologic de Telecomunicacions de Catalunya (CTTC), Castelldefels, Barcelona (Spain) `
Email: {jorge.baranda, pol.henarejos, ciprian.gavrincea}@cttc.es

**Abstract** — *This paper describes the implementation of a prototype of visible light communications system based on the IEEE 802.15.7 standard using low cost commercial off-the-shelf analog devices. The aim of this paper is to show that this standard provides a framework which could promote the introduction of applications into the market. Thus, these specifications could be further developed, reducing the gap between the industry and the research communities.*

*The implemented prototype makes use of Software Defined Radio platforms to interface between the analog devices and the computer where the signal processing is performed. The use of this concept provides the system with enough flexibility and modularity to include new features in the prototype without requiring long development time.*

## I. INTRODUCTION

During the last century, radio dominated the world of wireless communications. One could be surprised to find out that the first voice transmission over a wireless link has been done, not using radio, but using light waves. In 1880, Alexander Graham Bell had demonstrated the first wireless voice communications. Bell was able to clearly communicate over a distance of 213 m with his photophone, a device capable of transmitting data by modulating a beam of sunlight. Even though he had considered it as his greatest invention, the photophone was eclipsed by the later discovery of radio communications.

After the invention of mobile phone, the popularity of short range radio communications has increased. More and more mobile devices have appeared with an increasing need to exchange data wirelessly. This massive increase on the number of mobile devices is provoking a shortage in the radio spectrum resources. Under this scenario, the efforts of research community have been redirected towards the exploration of novel solutions that could guarantee an efficient usage of the available spectrum. As a consequence, new paradigms as cognitive radio, spectrum sharing and spectrum re-farming have emerged. This scenario, also offers a new opportunity for optical wireless communications (OWC) to regain the attention of wireless industry by providing additional spectrum resources.

OWC had a slow but constant evolution during the last century. Most of the OWC applications are related with short range and low data rate communication. They are using infra-red light and they can be found in our daily life, where the most known example is the remote control of electro domestic appliances. But, in the last years, visible light communication (VLC), a new paradigm of OWC has captured the attention of research community. As the name is suggesting, VLC is using beams of visible light to send the information. The main challenge of VLC systems consisted in finding a source of artificial light that can be easily modulated. The recent development of high power light-emitting diodes (LEDs) has provided a technical and cost effective solution for the aforementioned challenge.

The rapid evolution in the field of solid state electronics has opened the door to the development of lighting devices with better performance in terms of both energy efficiency and life expectancy whose cost is decreasing rapidly. Capitalizing on these properties, there is a growing number of application scenarios (hotels, hospitals, traffic lights, in-home applications), where this alternative is being used to replace incandescent light bulbs and fluorescent lamps. In fact, the market share of these devices is increasing year by year and it is expected that, by 2020, 70 percent on residential and 90 percent of architectural lighting will be LED-based [1].

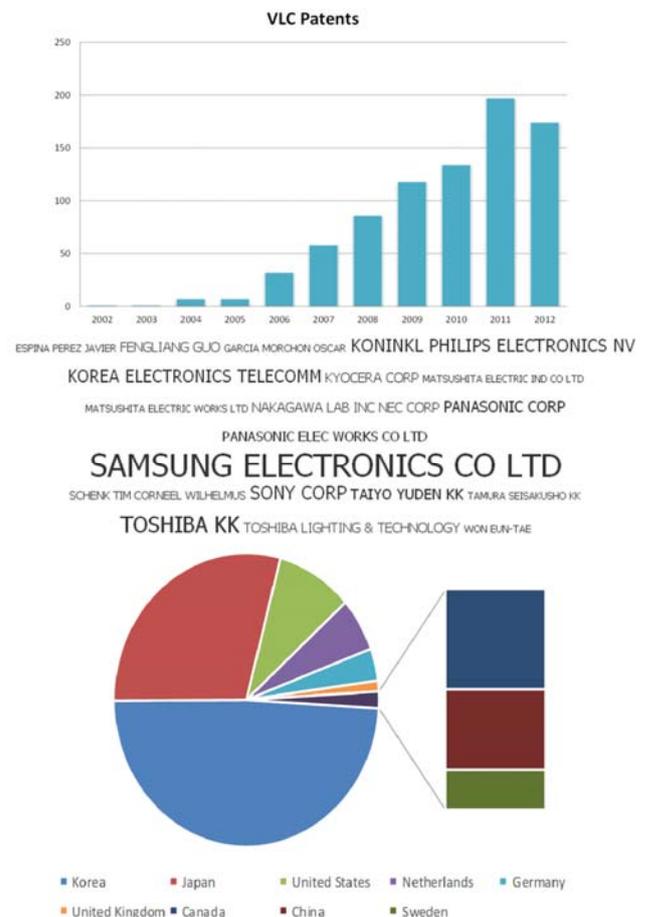

**Figure 1:** Evolution of VLC related patent applications. In the middle part, the principal applicants are listed. The font size is proportional with the number of patents. In the lower part, the geographical distribution of the patent number is presented.

As a matter of fact, these appliances might not serve only as illumination devices. During the last decade, the interest of communications research community in LED devices has been increasing because white-LEDs can be also used as data transmitters without losing their main functionality as illumination sources [2]-[3], enabling the appearance of VLC. This technology presents several advantages regarding radio-wave communications systems. Among them, robustness against electromagnetic interference and high level of protection against eavesdropping could be highlighted.

The research community is not the only one who gained interest in this new technology. The increasing number of patents related to VLC clearly demonstrates the interest of industry for this technology. A search conducted through the published patents, using the Patent Inspiration portal offered by CREAX Software NV reveals an increase in the number of patents related to VLC. A statistic of the VLC related patents, published during the last decade is illustrated in Figure 1. It can be noticed that the Asian industry (Samsung, Toshiba, Sony) has demonstrated the biggest interest in VLC.

The interest of academy and industry can also be illustrated by the publication of standard specifications for VLC. One standard for VLC has been published in 2007 in Japan [4] followed by a second standard published in 2011 by IEEE [5]. Nonetheless, although there are some standards already published for VLC communications there are not so many examples in the literature based on the utilization of such standards.

This paper presents system design guidelines for a software defined visible light communication (SD-VLC) transceiver compliant with IEEE 802.15.7 standard. They can serve for the development of new applications or standard based demonstrators which could promote the introduction of new applications into the market, awake the interest of industrial players and reduce the time-to-market figures for the development of future standard compliant products. The key implementation issues for a VLC system is to provide data communication capability to a LED luminary with minimum alteration of its illumination performance and production cost.

## II. VLC Systems Review

The interest shown by the research community in the field of VLC during the last years had led to the development of demonstrators capable of proving the feasibility of this novel technology for wireless applications. Based on the modulation technique used for transmitting the information, these demonstrators can be divided in two groups: one using binary-level modulation and the second one using multi-level modulation scheme.

Binary-level modulation refers to modulation techniques in which the information is sent in each symbol period through the variation of two intensity levels. Those techniques are popular schemes used mostly on wired communication and their main advantage is that they typically have simple and inexpensive implementations. An implementation based on Non-Return to Zero (NRZ) ON-OFF keying (OOK) offering a data rate of 40 Mb/s has been reported in [6]. The main limitation in case of binary-level modulation schemes is due to the small bandwidth offered by white-LED devices. To overcome this problem, a solution based on post-equalization has been proposed in [7] where data rate of 100 Mb/s for NRZ-OOK has been reported. The previous demonstrations have been done using opto-electronic receivers based on PIN photo-diode. Better results can be obtained by using avalanche photo-diodes (APD) in the construction of the opto-electronic receiver. In [8] a data rate of 230 Mb/s is reported. This performance is achieved by using OOK modulation and a VLC receiver based on APD.

Multi-level modulation refers to modulation techniques in which the information is sent by modifying the intensity values in a continuous range or by using a set of predefined values [9]. Because they provide a better usage of the available bandwidth, the systems based on this modulation schemes can achieve higher data rates. As a matter of fact, data rates of Gb/s are being reported in literature by using discrete multitone modulation (DMT). For example, in [10], VLC systems based on white-LED, providing data rates of 1.1 Gb/s are being presented. By using multicolor LED devices such as RGB LEDs, higher data rates can be obtained because multiple communication channels can be used. In [11], data rate of 3.4 Gb/s is being reported using RGB LED and DMT modulation scheme.

Analyzing the publications related to VLC it is clear that high data rates are achievable and that, placing this technology as a potential alternative to the one based on radio link. Taking a closer look at the experimental setup presented on those papers it can be seen that the results are obtained in special conditions and that in the majority of the cases the range of wireless link is on the order of ten of centimeters. Nevertheless, they represent great achievements and important proof that VLC can serve as a complementary technology for wireless communications.

In order to facilitate the entry of new technology on the market there is a need of applications and practical demonstrations that can prove the advantages offered by that technology. For the case of VLC, one of the main advantages consists on the dual functionality of illumination and data communications. The success of LED based illumination devices on the lighting market is due to their high power efficiency. Ideal for a VLC system will be to preserve this advantage and to use the success of LED lighting devices as an entry gate to the smart home/city market. Taking into account that the solutions based on binary-level modulation are power efficient while the ones based on multi-level modulation are bandwidth efficient, the binary-level modulation is preferred over the multi-level one. Moreover, practical implementations based on standard specifications are more appealing than custom made solutions when it comes to market penetration.

## III. IEEE 802.15.7 Physical Layer Overview

Aware of the potential of VLC in the future wireless networks, the research community together with important industrial partners decided to create the task group (TG7) inside the IEEE 802.15 working group, with the goal of providing a framework for defining VLC communications. As a result, a first release of IEEE 802.15.7 standard has been published in September of 2011 and presents rules for the implementation of VLC systems having two main functionalities: illumination and data communications.

In this specification, a physical (PHY) and media access control (MAC) layer for short-range optical wireless communications using visible light for indoor and outdoor applications are defined. Unlike other specifications such as[4], IEEE 802.15.7 pays special attention to problems specific to illumination systems, like flicker mitigation (related with eye-safety regulations) and dimming support (related to power savings and energy efficiency). This section provides a brief summary on the relevant aspects of the IEEE 802.15.7 standard [5][5].

*A. Modulation methods*

The IEEE 802.15.7 standard defines three physical layer types which are grouped by data rate. The supported data rates in each PHY were designed having in mind the possibility to support a wide range of optical sources and detectors. PHY I data rates range from 11.67 kb/s to 266.6 kb/s, PHY II operates from 1.25 Mb/s to 96 Mb/s, and PHY III operates from 12 Mb/s to 96 Mb/s. The modulation formats used in PHY I and PHY II systems are on-off keying and variable pulse position modulation (VPPM), which is a combination of pulse position modulation (2PPM) and pulse width modulation (PWM) for dimming support. Alternatively, PHY III uses a particular modulation format called color shift keying (CSK), where multiple optical sources are combined to produce white light.

**Table 1 PHY operating modes [5]**

| | | | PHY I OPERATING MODES | | |
|---|---|---|---|---|---|
| | | | FEC | | |
| Modulation | RLL code | Optical clock rate | Outer code (RS) | Inner code (CC) | Data rate |
| OOK | Manchester | 200 KHz | (15,7) | 1/4 | 11.67 kb/s |
| | | | (15,11) | 1/3 | 24.44 kb/s |
| | | | (15,11) | 2/3 | 48.89 kb/s |
| | | | (15,11) | none | 73.3 kb/s |
| | | | none | none | 100 kb/s |
| VPPM | 4B6B | 400 KHz | (15,2) | none | 35.56 kb/s |
| | | | (15,4) | none | 71.11 kb/s |
| | | | (15,7) | none | 124.4 kb/s |
| | | | none | none | 266.6 kb/s |

| | | PHY II OPERATING MODES | | |
|---|---|---|---|---|
| Modulation | RLL code | Optical clock rate | FEC | Data rate |
| VPPM | 4B6B | 3.75 MHz | RS (64,32) | 1.25 Mb/s |
| | | | RS(160,128) | 2 Mb/s |
| | | 7.5 MHz | RS(64,32) | 2.5 Mb/s |
| | | | RS(160,128) | 4 Mb/s |
| OOK | 8B10B | 15 MHz | none | 5 Mb/s |
| | | | RS(64,32) | 6 Mb/s |
| | | | RS(160,128) | 9.6 Mb/s |
| | | 30 MHz | RS(64,32) | 12 Mb/s |
| | | | RS(160,128) | 19.2 Mb/s |
| | | 60 MHz | RS(64,32) | 24 Mb/s |
| | | | RS(160,128) | 38.4 Mb/s |
| | | 120 MHz | RS(64,32) | 48 Mb/s |
| | | | RS(160,128) | 76.8 Mb/s |
| | | | none | 96 Mb/s |

As the scope of this paper is not focused on solutions using multiple optical sources, the information related to PHY III will not be presented.

OOK is a simple modulation in which a rectangular pulse is transmitted during a fixed time slot if the coded bit is '1', while the absence of the pulse during a time slot codifies the transmission of a '0'. PPM is also a modulation technique that uses rectangular pulses to code bits of information. In this case the width of the pulse, in time, is shorter than the size of the transmitting time slot. The position of the pulse inside the transmission time slot is used to code the information bits. For example, in the case of 2PPM, two pulses are used to code the information bits such that if the pulse is aligned with the beginning of the transmission time slot a '0' is transmitted while, if it is aligned with the end of transmission time slot, a '1' is transmitted. VPPM is a variation of 2PPM in which the width of the pulses can be modified, with the scope of controlling the power of the transmitted signal. A graphical representation of the two concepts of modulation is presented in Figure 2a.

The IEEE 802.15.7 standard is designed to work in several scenarios with the presence of optical noise sources such as natural daylight and artificial illumination devices. PHY I is optimized for low rate, long distance, outdoor applications such as vehicles, traffic lights and street lamps while PHY II is designed to work in indoor infrastructures and point-to-point applications using high data rates. Different forward error correction (FEC) schemes are included in each PHY layer definition, derived from the necessity to work in multiple scenarios and in the presence of hard decisions generated by the clock and data recovery (CDR) circuits. For outdoor applications, more resilient codes, e.g., concatenating Reed-Solomon (RS) and convolutional codes (CC), are used to counteract daylight interferences and larger path losses. Furthermore, an interleaving operation between the RS and the CC is inserted to increase the robustness. For indoor applications, where better channel propagation conditions are present, only RS codes are used for FEC, to achieve the proposed data rates.

Beside FEC procedures, run length limiting (RLL) coding is added to avoid potential intra-frame flickering and clock and data recovery detection problems derived from the presence of long runs of 1s and 0s. RLL codes guarantee DC balance at the output of the transmitter which guarantees a constant brightness level of the light source. Considered line codes in the IEEE 802.15.7 standard are Manchester, 4B6B, and 8B10B, which provide a tradeoff between overhead and implementation complexity. The different modulation schemes defined in the standard for the PHY I and PHY II layers are included in Table 1.

The optical clock rates chosen for the different PHY layers were also selected based on the application scenarios devised for each PHY type. The PHY I optical clock rate was chosen to be $\leq$ 400 KHz owing to the fact that transmitters used in such systems require high currents to drive the LEDs and hence the bandwidth of the device is limited. The PHY II optical clock rate is chosen to be $\leq$ 120 MHz because LEDs with better switching capabilities can be incorporated for indoor illumination or into mobile and portable devices.

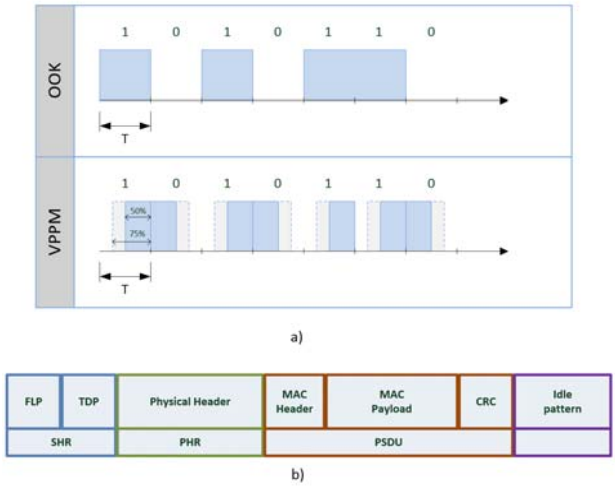

Figure 2: a) PHY I and PHY II modulation schemes
b) VLC frame structure [5]

*B. Frame format*

The frame defined by the IEEE 802.15.7 specification at physical level is constituted by three elements: the Synchronization Header (SHR), the Physical Header (PHR) and the Physical Service Data Unit (PSDU). The SHR contains the Fast Locking Pattern (FLP) and the Topology Dependent Pattern (TDP), which are needed to lock the CDR circuit and perform the synchronization with the incoming data flow. The PHR contains frame information related with the data length, the modulation and the FEC scheme used. The PSDU contains the MAC Header (MHR) where parameters of the MAC layer are included, such as the sequence number and control format fields. A graphical representation of the VLC frame structure is given in ¡Error! No se encuentra el origen de la referencia.b.

*C. Dimming methods*

Several mechanisms are included in the IEEE 802.15.7 standard to provide the system with dimming capabilities. Dimming is a feature present in current illumination systems which allows the user to control the brightness / dimming level of the light source. As the objective of the VLC systems is to provide data communication capabilities to the lighting infrastructure, it is important to preserve all the functionality and features offered by the current illumination systems.

For the case of OOK modulation, the dimming can be adjusted by means of two methods. The first one consists of redefining the "ON" and "OFF" levels to achieve the desirable brightness. The second method consists of varying the average duty cycle of the waveform by the insertion of "compensation symbols" into the modulation waveform. With this method, the part of the frame which contains the data bits is fragmented into sub-frames of the appropriate length and compensation symbols are added between the sub-frames with the scope of modifying the average brightness of the light source. The main disadvantage of this method is that the data rate decreases proportionally to the number of compensation symbols.

VPPM modulation allows dimming control due to its PWM characteristics. VPPM symbols '1' and '0' are distinguished by the pulse position within a unit time period. The dimming is adjusted changing the "ON" time pulse-width according to the requested dimming level. A dimming level resolution of 10% can be achieved by applying this technique.

It is important to remark that the dimming level shall be maintained independently on whether data is being sent or not. When data is not being transmitted, dimming can be adjusted by inserting idle patterns between data frames. The idle pattern can either be sent "in-band" or "out-of-band". The in-band idle pattern is sent at the same clock rate that the rest of the data frame and can be detected by the receiver. The out-of-band idle pattern is sent at a much lower optical clock rate (including the option of maintaining visibility via a DC bias only) and is not processed by the receiver.

## IV. SYSTEM DESIGN OF SOFTWARE DEFINED VLC SYSTEM

Rapid solutions for prototyping new standard compliant systems can play a key role in the integration of these standards into commercial applications. From a technical point of view, the most important properties of such prototype system are reprogrammability, reconfigurability and flexibility. From a commercial point of view, the main features are reduced cost and time to market figures. Reconfigurable hardware and solutions based on microprocessors or programmable logic devices are preferred for this type of implementations. Following this approach, robust prototyping systems can be obtained in a short period of time. The main disadvantage consist in the fact that highly specialized personal is required which makes difficult to fulfill the cost constraint.

Another approach for the implementation of prototyping systems is to use software techniques based on the concept of Software Defined Radio (SDR). The concept of SDR considers that the signal processing functions are performed in a general purpose processor (GPP), while the radio frequency (RF) and signal conversion (A/D, D/A) are performed in a programmable hardware. The fact that hardware problems are turned into software problems allows achieving a solution which requires less number of specialized personal and with a higher degree of modularity and flexibility than the one offered by HW solution.

According to the previous reasoning, the authors selected the SDR approach for developing a VLC prototype based on the IEEE 802.15.7. There is another implementation of a software defined VLC system [12], but to the best of our knowledge, the one herein presented represents the first implementation of a set of the physical layer features specified by the aforementioned standard. The aim of this section is to provide the reader with some guidelines to develop a low cost software defined VLC system using Commercial Off-The-Shelf (COTS) devices following the SDR approach.

The architecture of the proposed IEEE 802.15.7 prototype employing COTS devices is illustrated in Figure 3. The fact of using the SDR approach, divides the prototype into 2 subsystems: the hardware (HW) and the software (SW) subsystem. The hardware subsystem is formed by the opto-electronic devices, analog devices needed to control and to condition the signal originated or directed to the opto-electronic devices and data conversion modules. This is similar to SDR concept with the main difference being that the radio module is replaced by an opto-electronic module. The software subsystem is composed by two personal computers (PC), where the baseband signal processing is performed with the help of open source libraries.

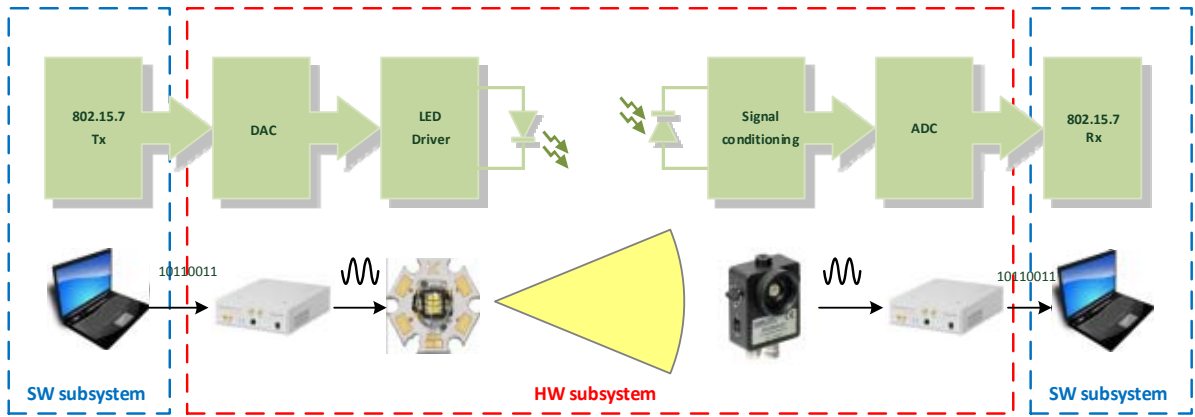

Figure 3: SD-VLC demonstrator block diagram

### A. Hardware subsystem

Typically, in software defined systems, the interface between HW and SW subsystems is done with the help of specialized devices which provide functionalities like data conversion and data buffering. In [13], a list of available commercial platforms can be found. From this list, the authors highlight the Universal Software Radio Peripheral (USRP) family of devices from Ettus Research because of its tradeoff between price and performance. These platforms are built around a field programmable gate array (FPGA) that includes powerful analog-to-digital and digital-to-analog converters and the possibility of adding extra conditioning circuitry with the help of extension boards.

A low cost solution for SD-VLC systems can be based on USRP platform, an amplification stage, the LED driver circuit and a commercial phosphorescent white LED as light source. The stream of bits generated at the transmitter software subsystem are delivered through a Gigabit Ethernet link to the USRP, where they are digital-to-analog converted. The modulated signal provided by the SDR platform has a low level voltage and has to be amplified in order to control the LED driver circuit. The light intensity generated by LED devices is proportional to the driving current and, as a consequence, the driver circuit should be able to control this current. A quick and easy solution that can be used for controlling high power LEDs consists of using a power MOSFET transistor, able to drive enough current to the emitting LED. Commercial LED luminaries are constructed combining high power LED device with optical components (e.g. reflectors, lenses) to provide a proper illumination according to the intended lighting application (room illumination, exhibition downlight). The light fixture of the VLC prototype presented in this paper has been designed using an OSTAR LE CW E2B LED device equipped with a reflector which reduces the viewing angle to 30º. Thus, the reflector shapes the light beam to concentrate it and overcome the problem of the broad viewing angle. The LED device has a typical luminous intensity of 90 cd and the resulting light fixture has a illuminance of 430 lux, measured at a distance of 1m.

At the receiver side, a photo-detector based on PIN or APD can be used. The signal provided by the photo-diode requires some conditioning processing. There are already commercial solutions which comes with the conditioning circuits incorporated into the device. One example is the PDA36A from ThorLabs, a compact photo-detector device equipped with a transimpedance amplifier with adjustable gain, ranging from 0 to 70 dB. The photo-detector delivers the received signal to the USRP receiving platform. There, the signal is sampled and passed to the receiving computer, where the demodulation is performed.

### B. Software subsystem

The modulation and the demodulation of the incoming bits to/from the data conversion platforms (USRP) are performed in a GPP by means of a signal processing open source library. An example of open source library for signal processing is GNU Radio. This library is an enormous body of pre-written, free software in continuous development by a community of programmers, who have developed blocks of code in C++ to handle a wide range of signal processing functions, as well as the blocks which interface the USRP devices and the GPP. Nevertheless, for this implementation, the authors developed a custom object-oriented framework combining GNU Radio with other well-known libraries such as QT library from Nokia. A detailed description of this platform can be found in [14]. This platform enhances the GNU Radio framework with a customized Graphical User Interface (GUI) which offers the possibility of choosing among a set of different communication system layouts. For each layout, a customized interface for system configuration and performance metric visualization is provided. Moreover, those libraries run over most operating systems (OS) offering the possibility of creating a platform independent solution. Figure 4 presents the baseband signal processing chains of the SD-VLC system. At the transmitter, the PHR and the PSDU units are generated independently according to the system configuration. Each unit passes through the PHY modulator block. The applied FEC scheme and RLL coding is selected with the GUI interface. After the modulation, both units are assembled into one stream of data and the TDP pattern is inserted. In the last block, before the USRP device, the FLP pattern and the in-band idle pattern are attached to constitute the entire frame.

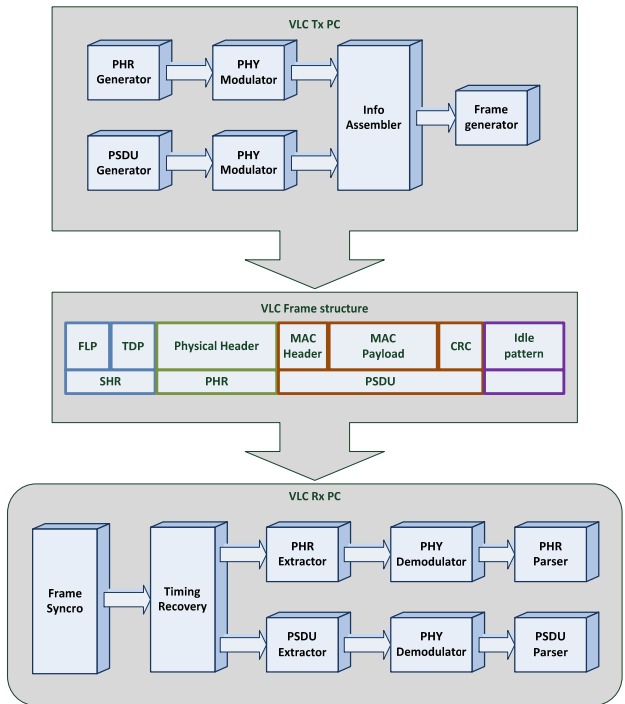

**Figure 4: PHY block diagram of SD-VLC system**

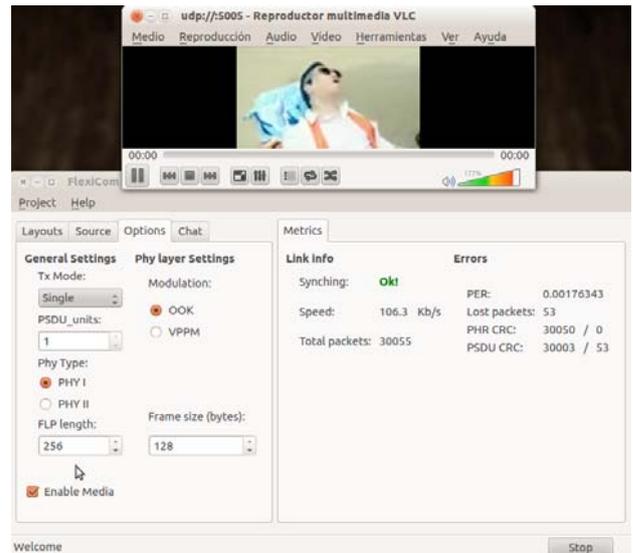

**Figure 5: Capture of receiver PC screen during video broadcasting demonstration**

At the receiver, frame synchronization and timing recovery are performed at the first block. Frame recovery is achieved through the exploitation of repeated patterns at the SHR header. Timing recovery is the procedure to determine the optimum sampling instant required to decide the value of the incoming bits. Timing recovery can be easily performed using a non-data aided detector based on the maximum likelihood (ML) algorithm. According to this detector, the optimum sample corresponds to the one that maximizes the energy of the received oversampled sequence of samples. Once the incoming flow of bits is synchronized, the PHR and the PSDU units are extracted and demodulated according to the system configuration. These units are parsed in order to extract its corresponding information and to check the validity of the received frame.

Currently, the SD-VLC prototype built following the previous guidelines supports all operating modes defined for PHY I, both for OOK and VPPM modulation methods. A more detailed description of the system and its performance are presented in [14]. Two multimedia applications have been developed to present the system capabilities to the general public. The first one, targeting a low data rate transmission, consists in a chat application which allows transmission of text messages between the two PCs of the VLC demonstrator. The second application, targeting the highest data rate transmission specified for PHY I, demonstrates the capability of streaming a compressed MPEG-TS video. Figure 5 presents the GUI interface developed using FlexiCom, a custom object-oriented framework for SDR applications developed at CTTC research centre, and some typical communications metrics captured during video broadcasting demonstration. Even though the transmission distance depends on the light intensity of the chosen LED device, it can be mentioned that the video transmission has been effectuated over a distance of 1.5 m while the chat application has been effectuated over a distance of 4 m.

## V. CONCLUSIONS

This paper presents the suitability of using the proposed IEEE 802.15.7 standard as a starting point to develop commercial low-medium data rate VLC applications. The herein presented proof-of-concept, combining software and hardware elements, offers a flexible platform where the introduction of new features can be achieved without requiring long development time. In conclusion, it can be said that the dual functionality requirement imposed to a VLC system can be easily achieved by following the IEEE 802.15.7 standard specifications. The low cost requirement imposed to a commercial VLC LED luminary can be fulfilled by porting the software define component of the presented solution to a commercial or specialized lighting microcontroller. The VLC literature presents many example of higher data rates transmission using various modulation schemes, but, as it was mentioned before, the objective of our system was to demonstrate and to present the capabilities of a standard compliant VLC system. The presented solution shows that the IEEE 802.15.7 specification is mature enough to deploy competitive solutions for the mentioned scenarios using COTS devices, not only in terms of costs but also in simplicity and performance.

VLC systems do not have to be seen as a substitutive technology for Wi-Fi or other wireless radio system, but as a complementary technology that will help to enhance the user experience and contribute to the development of the smart cities concept.

## REFERENCES

[1] McKinsey & Company, Lighting the way: Perspectives on the global lighting market, (July 2011).

[2] T. Komine, Y. Tanaka, S. Haruyama and M. Nakagawa, "Basic study on visible-Light communication using light emitting diode illumination", *Proc. Of 8th International Symposium on Microwave and Optical Technology*, pp. 45-48, 2001.

[3] T. Komine and M. Nakagawa, "Fundamental analysis for Visible-Light Communication System using LED Lights", *IEEE Transactions on Consumer Electronics*, vol. 50, no. 1, pp 100-107, Feb. 2004.

[4] Japan Electronics and Information Technology Industries Association (JEITA), Visible Light Communication CP-1221, CP-1222 specifications, Japan, 2007.

[5] Institute of Electrical and Electronics Engineers, "Standard for Local and metropolitan area networks, Part 15.7: Short-Range Wireless Optical Communication Using Visible Light", Rev. 802.15.7-2011, Sept. 2011.

[6] Le-Minh, D. O'Brien, G. Faulkner, L. Zeng, and K. Lee, 'High-Speed Visible Light Communications Using Multiple-Resonant Equalization', IEEE Photonics Technology Letters, 2008

[7] Le-Minh, D. O'Brien, at al. "100-Mb/s NRZ Visible Light Communications Using a Postequalized White LED", IEEE Photonics Technology Letters, 2009

[8] J. Vucic et al., "230 Mbit/s via a wireless visible-light link based on OOK modulation of phosphorescent white LEDs," in Proc. OFC/NFOEC, 2010, Paper OThH3.

[9] S. Hranilovic, "Wireless Optical Communication Systems", Springer Inc., New York, 2005

[10] F. M. Wu, C. T. Lin, at al. "1.1-Gb/s White-LED-Based Visible Light Communication Employing Carrier-Less Amplitude and Phase Modulation", IEEE Photonics Technology Letters, 2012

[11] G. Cossu, A. M. Khalid, P. Choudhury, R. Corsini, and E. Ciaramella, "3.4-Gb/s visible optical wireless transmission based on RGB LED," Optics Express ,2012.

[12] M. Rahaim, T. Borogovac, T.D.C. Little, A. Mirvakili and V. Joyner, "Demonstration of a Software Defined Visible Light Communication System", in Demo and Exhibits of International Conference on Mobile Computing and Networking, Las Vegas, USA, Sep. 2011.

[13] R. Farrell, M. Sanchez and G. Corley, "Software-Defined Radio Demonstrators: An Example and Future Trends",International Journal of Digital Multimedia Broadcasting, vol. 2009,

[14] J. Baranda, P. Henarejos, C. Gavrincea, " An SDR Implementation of a Visible Light Communication System based on IEEE 802.15.7 Standard", Proc. Of 20th International Conference on Telecommunications, ICT 2013,Casablanca, May 2013

**Ciprian George GAVRINCEA** is currently working as a senior researcher at Centre Tecnologic de Telecomunicacions de Catalunya (CTTC). Prior to his current position and until March 2008, he worked as lecturer at the North University of Baia Mare. He received his engineering degree in electro-mechanical engineering from the North University of Baia Mare in June 1999. Ten years later, he obtained his Ph. D. in electronics and telecommunications from the "Politehnica" University of Timisoara. He was involved in a number of industrial and public funded projects, being involved on the physical layer implementation of wireless communication systems like UWB and LTE. His research interests lie in the field of wireless communication, digital signal processing, embedded systems, artificial neural networks and recently on visible light communications.

**Jorge BARANDA** is currently working as a researcher at Centre Tecnologic de Telecomunicacions de Catalunya (CTTC). He received his MSc degree in communication technology from the Technical University of Catalonia UPC, Barcelona, Spain, in 2008. He has been involved in a number of industrial and public funded projects related with the development of physical layer implementation of wireless communication systems like WiMAX and LTE His research interests are focused on signal and image processing, radio wireless communication systems, efficient routing algorithms and Visible Light communication systems.

**Pol Henarejos** received the MSc. Telecommunication Engineering degree from the Technical University of Catalonia (UPC) in May 2009 and the European Master of Research on Information and Communication Technologies (MERIT) in 2012. He joined the CTTC in January 2010 as research engineer. During 2010, he participated in European projects implementing real receivers using the Filterbank Multicarrier approach. After that, he was involved in industrial projects based in implementations of the physical layer aspects of LTE in prototypes. Such skills permit to deploy the new standard for interactive services in satellite communications and test innovative techniques. His interests comprise from the implementations of physical layer of radio communications in real devices to theoretical studies in resource management in multiuser scenarios.